\documentclass[conference]{IEEEtran}
\usepackage{enumitem}
\usepackage{graphicx,multicol} 
\usepackage{epsfig} 
\usepackage{amsmath,amsthm} 
\usepackage{amssymb}  
\usepackage{float}
\usepackage{cases,setspace,adjustbox}
\usepackage{cite}
\usepackage{algorithmic}
\usepackage[ruled,vlined]{algorithm2e}
\usepackage{array}
\usepackage[update,prepend]{epstopdf}
\usepackage{eqparbox}
\usepackage[font=footnotesize]{subfig}
\usepackage{fixltx2e}
\usepackage{url}
\usepackage{amsthm}
\usepackage{array}
\usepackage{caption}
\usepackage{xcolor}
\newcommand{\pidlenet}{p_\text{I}}
\newcommand{\pidle}[1]{p_\text{I}^{(#1)}}
\newcommand{\psuccnet}{p_\text{S}}
\newcommand{\psucc}[1]{p_\text{S}^{(#1)}}
\newcommand{\tauD}{\tau_{D}}
\newcommand{\tauW}{\tau_{W}}
\newcommand{\NED}{\widehat{\tauD}}
\newcommand{\NEW}{\widehat{\tauW}}
\newcommand{\SED}{\widetilde{\tauD}}
\newcommand{\SEW}{\widetilde{\tauW}}
\newcommand{\OPTD}{\tauD^{*}}
\newcommand{\OPTW}{\tauW^{*}}
\newcommand{\SG}[1]{{\color{black} {#1}}}
\newcommand{\SnG}[1]{{\color{black} {#1}}}
\newtheorem{lem}{Lemma}
\newtheorem{theorem}{Theorem}
\begin{document}
\renewcommand{\arraystretch}{1.15}
\title{\huge A Game Theoretic Approach to DSRC and WiFi Coexistence}
\date{}
\author{Sneihil Gopal and Sanjit K. Kaul\\
Wireless Systems Lab, IIIT-Delhi, India\\
\{sneihilg, skkaul\}@iiitd.ac.in}

\maketitle
\begin{abstract}
We model the coexistence of DSRC and WiFi networks as a strategic form game with the networks as the players. Nodes in a DSRC network must support messaging of status updates that are time sensitive. Such nodes would like to achieve a small age of information of status updates. In contrast, nodes in a WiFi network would like to achieve large throughputs. Each network chooses a medium access probability to be used by all its nodes. We investigate Nash and Stackelberg equilibrium strategies.
\end{abstract}

\IEEEpeerreviewmaketitle
\section{Introduction}
The Federal Communications Commission (FCC) had previously allocated the $5.9$ GHz band for vehicle-to-vehicle (V2V) or vehicle-to/from-infrastructure (V2I) communications, also known as Dedicated Short Range Communications (DSRC). These communications include that of safety messaging, which involves periodic broadcast of time sensitive state information by vehicles. Also, they use the physical and MAC layer as described in the IEEE $802.11$p standard, which like WiFi uses a \SnG{carrier sense multiple access and collision avoidance} (CSMA/CA) based medium access mechanism.

\begin{figure}[t]
\begin{center}
\includegraphics[width=0.75\columnwidth]{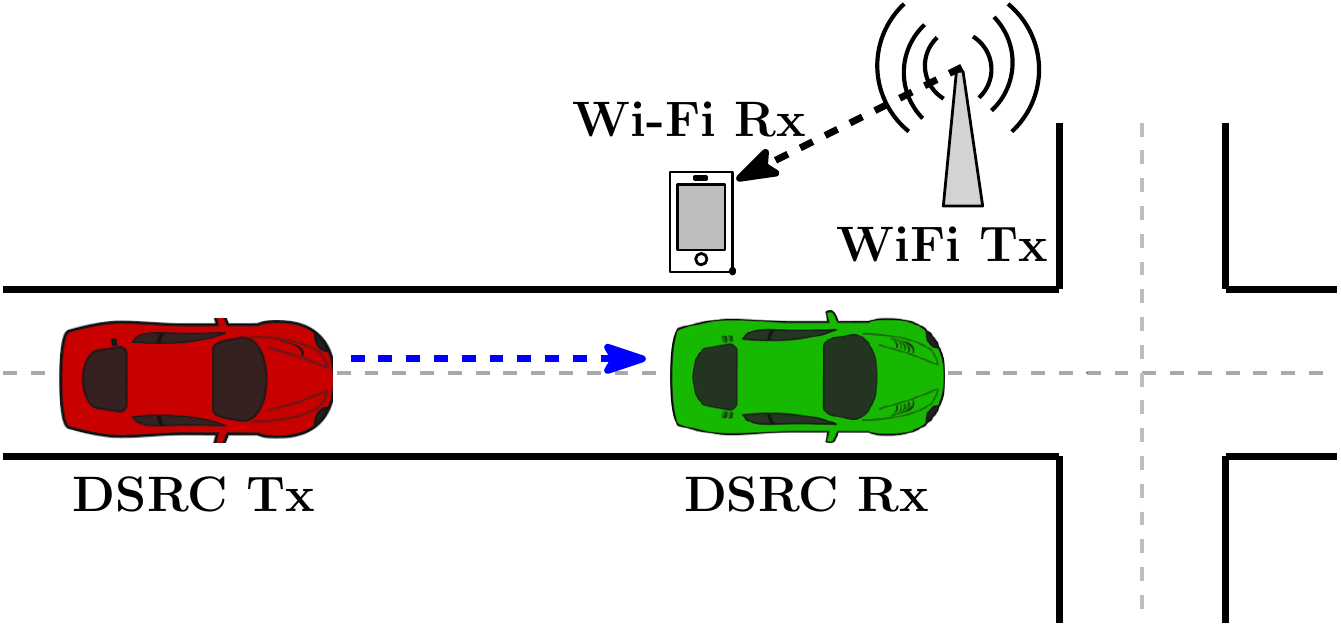}
\caption{\small Illustration of a vehicle-to-vehicle link in \SG{the} vicinity of a WiFi network.}
\label{fig:example}%
\vspace{-1em}
\end{center}
\end{figure}

More recently, to better support the demand for high data rates, the FCC opened up $195$ MHz of additional spectrum for use by unlicensed devices in the $5.35-5.47$ GHz and $5.85-5.925$ GHz bands. This additional spectrum can allow $802.11$ac based WiFi networks to accommodate additional wide-bandwidth channels and therefore enhance support for high data-rate applications. However, the $5.85-5.925$ GHz band overlaps with the band reserved for vehicular communications leading to the possibility of WiFi networks deployed in this band interfering with V2V and V2I communications. Figure~\ref{fig:example} \SnG{provides an illustration} in which a WiFi access point communicates with its client in the vicinity of a V2V link.

The resulting network coexistence problem is interesting as while both WiFi and vehicular communications use similar medium access mechanisms they aim to optimize very different utilities. While WiFi networks would like to maximize throughput, vehicular safety messaging is contingent on timely delivery of state information being broadcast by vehicles. This requirement of timely delivery of state is captured well by the metric of the age of information~\cite{kaul2012real}. 

In this work, we investigate a game theoretic approach to the coexistence of DSRC and WiFi, where while the DSRC network aims to minimize the age of information of updates, WiFi devices seek to maximize their throughput. Our specific contributions are as follows:
\begin{itemize}
\item We formulate a network coexistence game in strategic form, with DSRC and WiFi networks as players. The DSRC network \SnG{desires a small average age} and the WiFi network \SnG{desires a large average} throughput. The networks pay a cost for transmission opportunities that are wasted because of no transmissions by nodes or because of collisions (multiple transmissions in the same slot). Each network chooses a probability of medium access for all of its nodes. 

\item We investigate Nash equilibrium (NE) and Stackelberg equilibrium (SE) strategies. We demonstrate the efficacy of these strategies via simulations.
\end{itemize}

The rest of the paper is organized as follows. Section~\ref{sec:related} describes the related works. The network model is described in Section~\ref{sec:model}. This is followed by formulation of the game in Section~\ref{sec:game}. The Nash and Stackelberg strategies are described in Section~\ref{sec:nash} and~\ref{sec:stackelberg}, respectively. In Section~\ref{sec:results} we discuss numerical results obtained from example networks. We end with a summary of our observations in Section~\ref{sec:conclusion}.
\section{Related Work}
\label{sec:related}
Recent works such as~\cite{lansford2013,liu2017,naik2017,khan2017} study the coexistence of DSRC and WiFi. In~\cite{lansford2013} authors provide an overview of $5.9$ GHz band sharing. In~\cite{liu2017} and~\cite{naik2017} authors study the impact of DSRC on WiFi and vice versa. In these earlier works authors provide an in-depth study of the inherent differences between the two technologies, the coexistence challenges and propose solutions to better coexistence. In our work, we look at the coexistence problem as that of coexistence of networks that use similar access mechanism but have different objectives.

Works such as~\cite{sun2017update,he2016optimizing,he2017optimal,sun2016optimizing,yates2017status,kaul2011minimizing} investigate age of information in wireless networks. In~\cite{sun2017update} authors study optimal control of status updates from a source to a remote monitor via a network server. In~\cite{he2016optimizing} and~\cite{he2017optimal}, authors investigate scheduling strategies that minimize age. In~\cite{sun2016optimizing}, authors study age optimization without loss in throughput. In~\cite{yates2017status}, authors analyze the age of information over multiaccess channels. In~\cite{kaul2011minimizing}, authors investigate minimizing the age of status updates sent by vehicles over a carrier-sense multiple access (CSMA) network.

In this paper, we propose a game theoretic approach to study the coexistence of DSRC and WiFi. While throughput performance as the payoff function has been extensively studied from the game theoretic point of view (for example, see~\cite{mario2005,chen2010}), age of information as a payoff function has not garnered much attention yet. In a recent work~\cite{impact2017}, authors formulate a two player game where one player aims to maintain the freshness of information updates while the other player aims to prevent this. 
\section{Network Model}
\label{sec:model}
Our network consists of $N_D$ DSRC and $N_W$ WiFi nodes that contend for access to the shared wireless medium. Let $\mathcal{N}$ be the set of all nodes in the network. In practice, both DSRC and WiFi nodes access the medium using a binary exponential backoff (BEB)~\cite{bianchi} based CSMA/CA mechanism. While a WiFi network typically uses multiple backoff stages, the DSRC network uses just one backoff stage for safety messaging~\cite{liu2017}. 

In this work, we eschew the detailed workings of the BEB protocol. Instead we approximate its workings in the following manner. We assume that all nodes can sense each other's packet transmissions. This allows modeling the BEB as a slotted multiaccess system. Further, we assume that all nodes always have a packet to send. Also, a packet transmission by a node has a constant probability of collision. These latter two assumptions allow summarizing the parameters of the BEB (number of backoff stages and contention window sizes~\cite{bianchi}) in terms of the probability with which a node transmits in a slot and the probability that a node's transmission will end in a collision. 

Let $\tau_i$ denote the access probability with which node $i$ accesses the wireless medium. Let $\pidlenet$ be the probability of an idle slot, which is a slot in which no node transmits. Also, let $\pidle{-i}$ be the probability of the event that no node in the set of nodes not including node $i$, transmits in a slot. We have
\begin{align}
\pidlenet = \prod\limits_{i=1}^{|\mathcal{N}|} (1-\tau_i)\text{ and }
\pidle{-i} = \prod\limits_{\substack{j=1\\j \neq i}}^{|\mathcal{N}|} (1-\tau_j).
\label{eqn:idleprobs}
\end{align}
A packet transmission by a node is successful in a slot only if no other node transmits packet in the slot. Let $\psuccnet$ be the probability of a successful transmission in a slot, $\psucc{i}$ be the probability of a successful transmission by node $i$, and $\psucc{-i}$ be the probability of a successful transmission in a slot by a node other than node $i$. We have
\begin{align}
\psuccnet&= \sum\limits_{i=1}^{|\mathcal{N}|}\tau_i\prod\limits_{\substack{j=1\\j\neq i}}^{|\mathcal{N}|} (1-\tau_j),\quad
\psucc{i}= \tau_i\prod\limits_{\substack{j=1\\j\neq i}}^{|\mathcal{N}|} (1-\tau_j),\nonumber\\
&\qquad\text{ and }\psucc{-i}= \sum\limits_{\substack{j=1\\j\neq i}}^{|\mathcal{N}|}\tau_j\prod\limits_{\substack{k=1\\k\neq j}}^{|\mathcal{N}|}(1-\tau_k).
\label{eqn:succprobs}
\end{align}
Let $\sigma_I, \sigma_S$ and $\sigma_C$ denote the lengths of an idle, successful and collision slot, respectively.

Next we define the throughput of a WiFi node and the age of information (AoI) of a DSRC node in terms of the above probabilities and slot lengths.

\subsection{Throughput of a WiFi node}
Given the assumption that a node always has a packet to transmit, we may define the throughput of a node~\cite{bianchi} to be the fraction of an average slot that constitutes a successful transmission by the node. Let $T_i$ be the throughput of WiFi node $i$. We define
\begin{align}
T_i = \frac{\psucc{i}\sigma_S}{\sigma_I \pidlenet + \sigma_S \psuccnet + \sigma_C (1-\pidlenet-\psuccnet)}.\label{Eq:Throughput}
\end{align}
Note that the denominator in the above equation is the expected value of the length of a slot. The numerator is \SnG{the} average length occupied by a successful transmission by $i$.
\subsection{Age of Information of a DSRC node}
\begin{figure}[t!]
\begin{center}
\includegraphics[width=1\columnwidth]{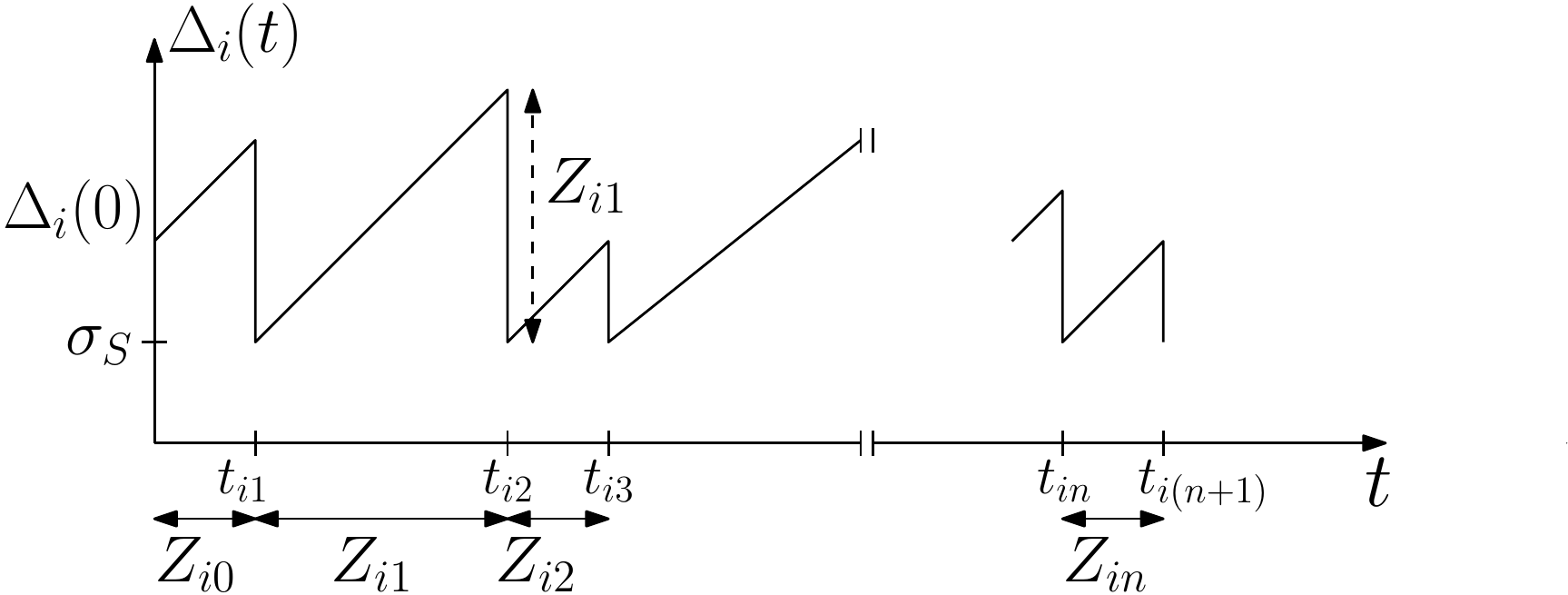}
\caption{Example sample path of the age $\Delta_{i}(t)$ of DSRC node $i$. The time $t_{ij}$ is the time of its $j^{th}$ successful update.}
\label{fig:CSMA}%
\end{center}
\vspace{-1em}
\end{figure}
Figure~\ref{fig:CSMA} shows an example sample path of the instantaneous age $\Delta_{i}(t)$ of a certain DSRC node $i$. We will calculate the age of information of node $i$ using time average analysis as in~\cite{yates2017status}. We summarize it here for completeness.
 
Let $t_{ij}$ be the time of the $j^{th}$ successfully transmitted update packet by node $i$. The time between the $j^{th}$ and $(j+1)^{th}$ update by node $i$ is the inter-update time $Z_{ij} = t_{i(j+1)} - t_{ij}$. Let the $Z_{ij}$ of node $i$ be identically distributed as $Z_i$. The time average age of updates of $i$ is the area under the age graph normalized by the time interval of observation. Over an interval $(0,t')$, it is $\langle\Delta_i\rangle_{t'} = \frac{1}{t'}\int_{0}^{t'} \Delta_{i}(t) dt.$ We can rewrite $\langle\Delta_i\rangle_{t'}$ as
\begin{align}
&\langle\Delta_i\rangle_{t'} = \frac{T_{0}}{t'}+\frac{n}{t'}\frac{1}{n}\sum\limits_{j=1}^{n}\left(\frac{Z_{ij}^2}{2} + \sigma_S Z_{ij}\right).
\end{align}
where, $T_{0} = \frac{Z_{i0}^{2}}{2} + Z_{i0}\Delta_{i}(0)$ and $\sigma_S$, \SnG{as before, is the length of successful slot, which is the time a successful update takes to get transmitted}. The age of information (AoI) \SnG{of} DSRC node $i$ is
\begin{align}
&\Delta_{i} = \lim\limits_{t'\rightarrow \infty}\langle\Delta_i\rangle_{t'} = \frac{E[Z_{i}^2]}{2E[Z_i]}+\sigma_S. \label{eq:AoI}
\end{align}

Next we characterize $Z_i$ in terms of the slot lengths and the probabilities in equations~(\ref{eqn:idleprobs}) and~(\ref{eqn:succprobs}).

The inter-arrival time $Z_i$ consists of a random number $L$ slots. The first $L-1$ of these slots are slots in which node $i$ is unable to transmit successfully. This is followed by the $L^{th}$ slot in which $i$ transmits successfully. We can write the inter-update time $Z_i$ as
\begin{align}
Z_{i} = Y_1+ Y_2+\dots+Y_{L-1}+ X.
\end{align}
The random variable $X$ is the length of $L^{th}$ slot and takes a value of $\sigma_s$ with probability $1$. Now consider $Y_l$, which is the length of $l^{th}$ slot, where $1\leq l \leq L-1$. In such a slot node $i$ does not update successfully, which is an event that occurs with probability $1-\psucc{i}$. Such a slot could therefore be an idle slot, or a slot that has a successful transmission by a node other than the node $i$, or a slot that has a collision. The PMF of $Y_l$ is thus given by
\begin{align}
    P[Y_l=y] =
    \begin{cases}
      \frac{\pidlenet}{1-\psucc{i}} & y=\sigma_I, \\
      \frac{\psucc{-i}}{1-\psucc{i}} & y=\sigma_S,\\
      \frac{1-\pidle{-i}-\psucc{-i}}{1-\psucc{i}} & y=\sigma_C,\\
      0 & \text{otherwise}.
    \end{cases}
    \label{eqn:pmfYl}
\end{align}

The number of slots $L$ is a geometric random variable with PMF
\begin{align}
    P[L=l] =
    \begin{cases}
      \psucc{i}(1-\psucc{i})^{l-1} &  l\geq 1, \\
      0 & \text{otherwise}.
    \end{cases}
\label{eqn:pmfL}  
\end{align}
Note that the random variables $Y_l$, $1\le l \le L-1$, $X$, and $L$ are mutually independent. Using this fact, we can write
\begin{align}
E[Z_{i}] &= (E[L]-1) E[Y] + E[X],\label{eq:firstmoment}\\
E[Z_i^2]&=(E[L]-1)(E[Y^2]+2E[X]E[Y])\nonumber\\
&\quad + (E[L^2]-3E[L]+2)E[Y]^2 + E[X^2]. \label{eq:secondmoment}
\end{align}

Note that $E[X]=\sigma_S$ and $E[X^2]=\sigma_S^2$. The moments of $Y$ and $L$ can be obtained from their PMF(s) given by equations~(\ref{eqn:pmfYl}) and~(\ref{eqn:pmfL}), respectively.   They can be used to calculate the moments of $Z_i$, which can be used to calculate AoI by using~(\ref{eq:AoI}). Lemma~\ref{thm:csmaAoI} summarizes these moments and the resulting AoI.

\begin{lem}
For CSMA/CA based access, the first and second moments of the inter-update time of node $i$ and the corresponding age of information are given by
\begin{align}
E[Z_i] &= \frac{\sigma_I \pidlenet + \sigma_S (\psucc{-i} + \psucc{i}) + \sigma_C (1-\pidle{-i}-\psucc{-i})}{\psucc{i}},\nonumber\\
E[Z_i^2]&= 2E[Z_i]^2 +\sigma_S^{2} - 2\sigma_S E[Z_i]\nonumber\\
&+\frac{\sigma_I^2 \pidlenet + \sigma_S^2 \psucc{-i} + \sigma_C^{2} (1-\pidle{-i}-\psucc{-i})}{\psucc{i}},\nonumber\\
\Delta_i &= \frac{\sigma_I \pidlenet + \sigma_S (\psucc{-i}+\psucc{i})+\sigma_C(1-\pidle{-i}-\psucc{-i})}{\psucc{i}}\nonumber\\
&+ \frac{\sigma_I^2 \pidlenet+\sigma_S^2(\psucc{-i}+\psucc{i})+\sigma_C^{2}(1-\pidle{-i}-\psucc{-i})}{2(\sigma_I\pidlenet+\sigma_S (\psucc{-i}+\psucc{i})+\sigma_C(1-\pidle{-i}-\psucc{-i})}.\nonumber
\end{align}
\label{thm:csmaAoI} 
\end{lem}
\section{Game Model}
\label{sec:game}
We model the coexistence of DSRC and WiFi networks as a non-cooperative game. This network coexistence game is defined in strategic form next. We have
\begin{itemize}
\item \textbf{Players:} We have two players namely the DSRC and WiFi networks. Recall they have $N_D$ and $N_W$ nodes, respectively.
\item \textbf{Strategy:} Each network can choose the probability with which all its nodes access the shared medium from $(0,1)$. Let $\tau_D$ and $\tau_W$ be the choice of access probabilities made by the DSRC and WiFi networks, respectively.
\item \textbf{Payoff:} The DSRC network \SnG{has an} average age of information $\Delta(\tau_D,\tau_W)$. On the other hand, the WiFi network \SnG{achieves an} average throughput $T(\tau_D,\tau_W)$. Let $c(\tau_D,\tau_W)$ be the cost of playing the game. 
The payoffs of the DSRC and WiFi networks are respectively
\begin{align}
u_D(\tau_D,\tau_W) &= -\Delta(\tau_D,\tau_W) - c(\tau_D,\tau_W),\label{eq:agepayoff}\\
u_W(\tau_D,\tau_W) &= T(\tau_D,\tau_W) - c(\tau_D,\tau_W).\label{eq:thrpayoff}
\end{align}
The networks would like to maximize their payoffs.
\end{itemize}

In what follows, we will set $\sigma_I = \beta$, $0 < \beta < 1$, and we will assume that the lengths of a successful transmission slot and that of a collision slot are the same. Specifically, $\sigma_S = \sigma_C = (1+\beta)$. \SnG{In practice, the idle slot is much smaller than a collision or a successful transmission slot, that is, $\beta << 1$.} The resulting average throughput and average age of information are
\begin{align}
T(\tauD,\tauW)& = \frac{\tauW(1-\tauW)^{(N_W-1)}(1-\tauD)^{N_D}(1+\beta)}{1-(1-\tauW)^{N_W}(1-\tauD)^{N_D}+\beta},\label{Eq:thrutility}\\
\Delta(\tauD,\tauW)& = \frac{1-(1-\tauD)^{N_D}(1-\tauW)^{N_W}+\beta}{\tauD(1-\tauD)^{(N_D-1)}(1-\tauW)^{N_W}} + \frac{\beta}{2}\nonumber\\  &+\frac{(1+\beta)(1-(1-\tauD)^{N_D}(1-\tauW)^{N_W})}{2(1-(1-\tauD)^{N_D}(1-\tauW)^{N_W}+\beta)}.\label{Eq:ageutility}
\end{align}

\subsection{Cost of idling and collision}
We penalize the networks for wastage of spectrum that occurs as a result of idle slots or due to collisions (say because either network chooses an aggressive access probability). The cost function $c(\tau_D,\tau_W)$ is defined as
\begin{align}
c(\tau_D,\tau_W) &= w_{idle}\pidlenet + w_{col}(1-\pidlenet-\psuccnet)\\
&= w_{idle}(1-\tauD)^{N_D}(1-\tauW)^{N_W}\nonumber\\
&\quad+ w_{col}\Bigl(1-(1-\tauD)^{N_D}(1-\tauW)^{N_W}\nonumber\\
&\quad -\SG{N_D}\tauD(1-\tauD)^{(N_D-1)}(1-\tauW)^{N_W} \nonumber\\
&\quad -\SG{N_W}\tauW(1-\tauW)^{(N_W-1)}(1-\tauD)^{N_D}\Bigr)
\label{eq:cost}
\end{align}
where, \SG{$w_{idle}\geq 0$ and $w_{col}\geq 0$} allow flexibility in how idle and collision slots are penalized, for example, by a spectrum regulator, with the goal of enabling effective use of the shared spectrum.  

Next we consider in turn the Nash equilibrium strategy ($\NED,\NEW$) and the Stackelberg strategy ($\SED,\SEW$).
\subsection{Nash Strategies}
\label{sec:nash}
\SG{As stated in~\cite{zuhan}, a pure strategy Nash equilibrium of a non-cooperative game $G = (\mathcal{N},(S_i)_{i\in\mathcal{N}},(u_i)_{i\in\mathcal{N}})$ is a strategy profile $\mathbf{s^{*}}\in S$ such that $\forall i \in \mathcal{N}$
\begin{align}
u_{i}(s_{i}^{*},\mathbf{s_{-i}^{*}}) \geq u_{i}(s_{i},\mathbf{s_{-i}^{*}}), \quad \forall s_{i} \in S_i,
\end{align}
\SnG{where $\mathcal{N}$ is the set of players, $S_i$ is the strategy set of player $i$ and $u_i$ is the payoff of $i$.} 

Also, a non-cooperative game $G$ has atleast one pure strategy Nash equilbrium if $\forall i\in \mathcal{N}$ 
\begin{enumerate}
\item the corresponding strategy set $S_i$ is non-empty, convex and compact subset of some Euclidean space $\mathbb{R}^M$ 
\item $u_{i}(s_i,\mathbf{s_{-i}})$ is a continuous function in the profile of strategies $\mathbf{s}\in S$ and quasi-concave in $s_i$.
\end{enumerate}
The following theorem states the existence of NE for our network coexistence game.
\begin{theorem}\label{thm:quasiConcavity}
The network coexistence game defined in Section~\ref{sec:game} has at least one pure strategy Nash Equilibrium. Specifically, the payoffs $u_D(\tau_D,\tau_W)$ and $u_W(\tau_D,\tau_W)$ are quasi-concave in $\tau_D\in (0,1)$ and $\tau_W\in (0,1)$, respectively, for $w_{idle}, w_{col} \ge 0$, $0 < \beta < 1$, and $N_D, N_W \ge 1$.
\end{theorem}
The proof is given in the Appendices at the end.} 

%

\subsection{Stackelberg Strategies}
\label{sec:stackelberg}
We study the Stackelberg equilibrium when the DSRC network is the leader and the WiFi network is the follower and vice versa. We first discuss the game where the DSRC network is the leader and define the optimal response (or optimal reaction) set $R_{W}$ of the follower (the WiFi network) to the strategy of the leader. The optimal response of the WiFi network is given by
\begin{align}
R_W(\tauD) = \{\tauW \in (0,1): u_W(\tauD,\tauW)\geq u_W(\tauD,t),\nonumber\\
\forall t \in (0,1)\}.\label{Eq:bestresponse}
\end{align}
We now define the \textit{Stackelberg equilibrium strategy} for the DSRC network. A strategy $\SED$ is called a Stackelberg strategy for the leader, if
\begin{align}
\min_{\tauW\in R_W(\SED)} u_D(\SED,\tauW) =  \max_{\tauD\in(0,1)} \min_{\tauW\in R_W(\tauD)}u_D(\tauD,\tauW),
\label{Eq:SE}
\end{align}
where $\min_{\tauW\in R_W(\SED)} u_D(\SED,\tauW)$ is the Stackelberg payoff for the leader. The definition for when the WiFi network is the leader can be obtained by swapping the subscripts $D$ and $W$ in equations~(\ref{Eq:bestresponse}) and~(\ref{Eq:SE}).

\section{Evaluation Methodology and Results}
\label{sec:results}
In this section, we discuss results from simulations. We begin by discussing the impact of access probability on the average throughput and average AoI. We compute the Nash equilibrium strategies for different selections of $N_D$ and $N_W$ and highlight the impact of selection of weights $w_{idle}$ and $w_{col}$ on the Nash equilibrium strategies and the obtained AoI and throughput values. Finally, we compare the NE strategies with the Stackelberg equilibrium strategies\footnote{To solve for the equilibrium, we linearly rescale $\Delta$ to the range of $T$.}. For the results shown, the networks were allowed to choose access probabilities $\tau_D$ and $\tau_W$ from the interval $[0.01,0.99]$.

\begin{figure}[t!]
\vspace{-15pt}
  \centering
  \subfloat[$N_D = 1$]{\includegraphics[width=0.48\columnwidth]{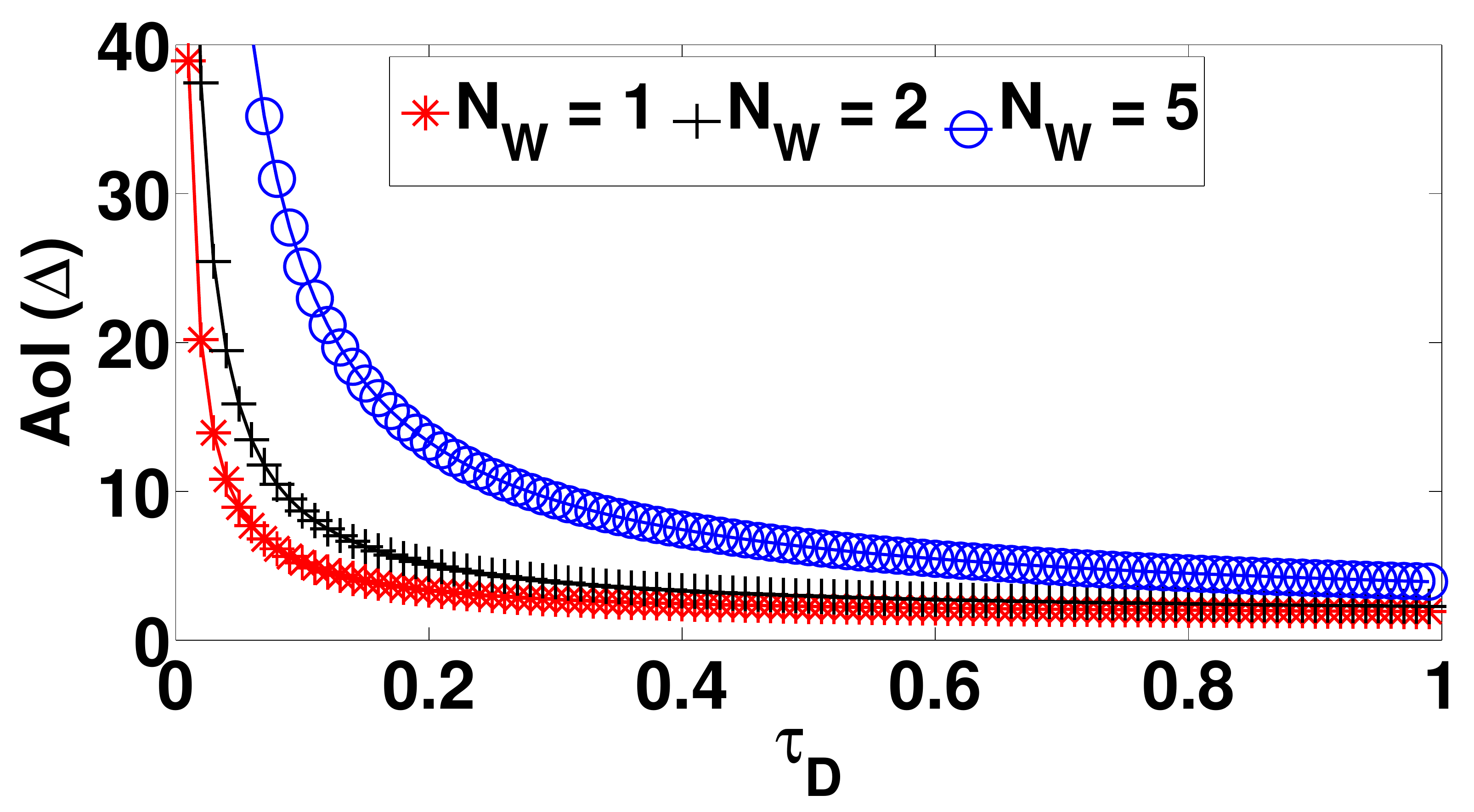}\label{fig:Age1}}
  \enspace
  \subfloat[$N_D = 5$]{\includegraphics[width=0.5\columnwidth]{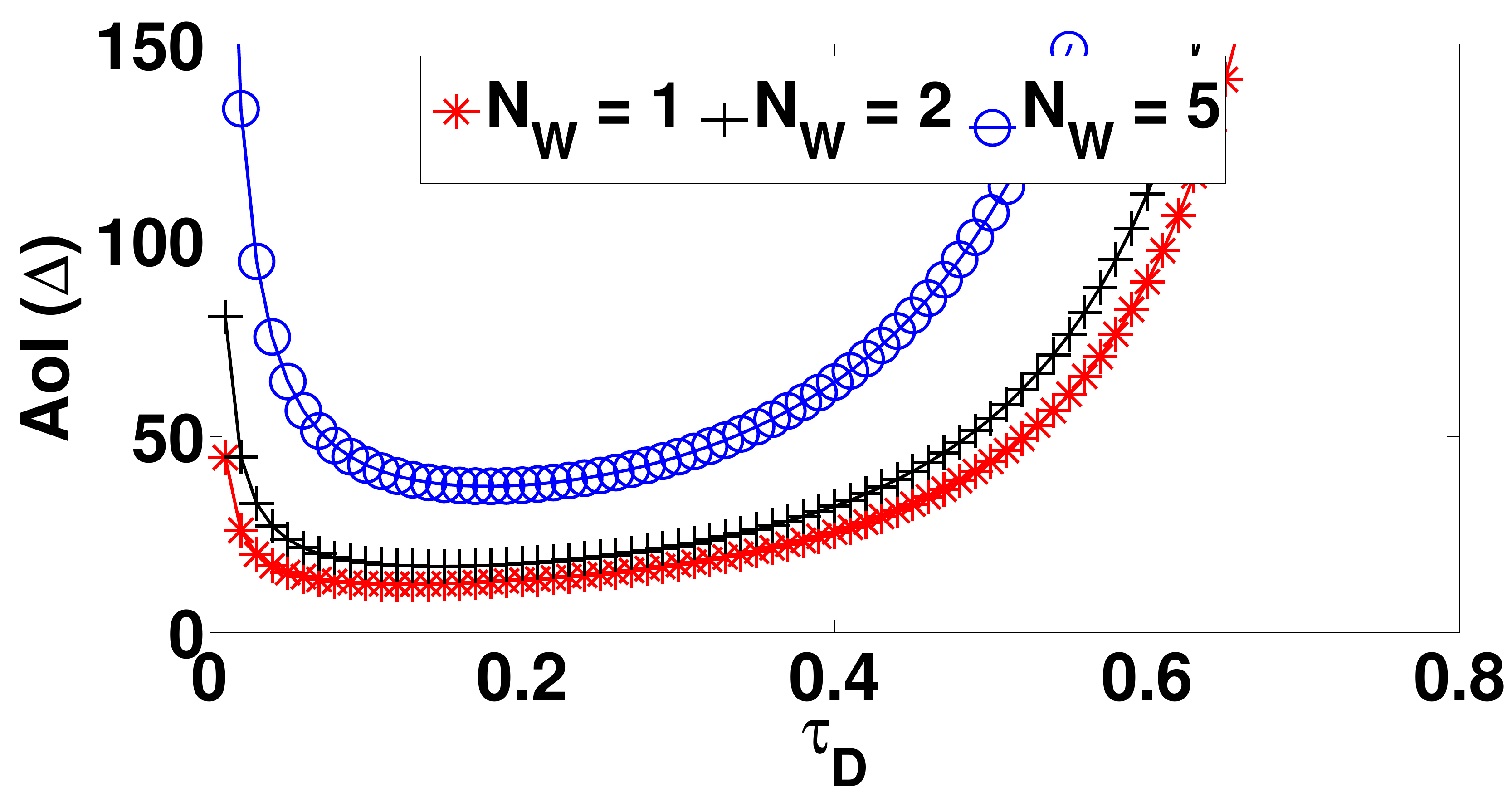}\label{fig:Age3}}\\
  \subfloat[$N_W = 1$]{\includegraphics[width=0.48\columnwidth]{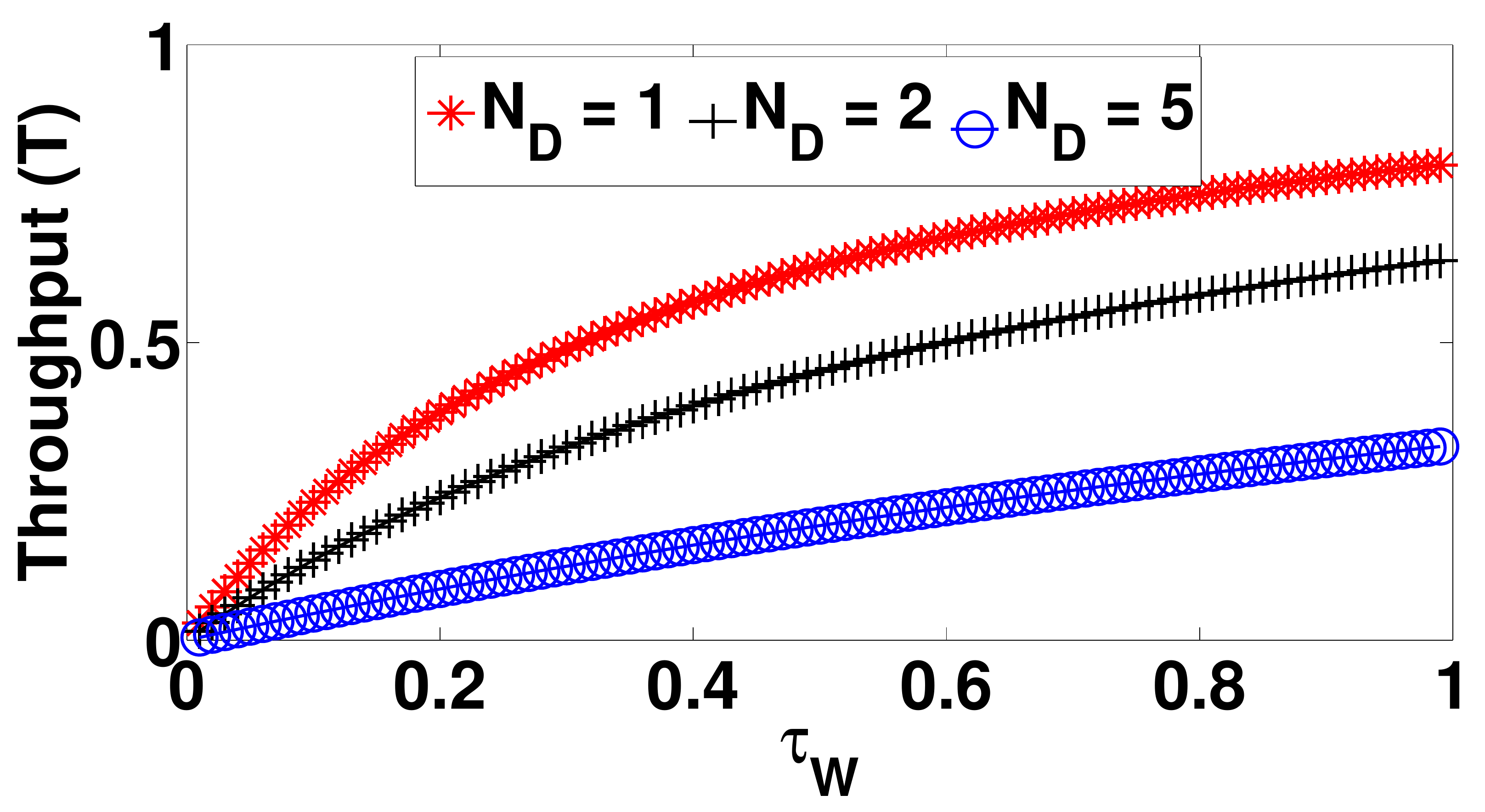}\label{fig:Thr1}}
  \enspace
  \subfloat[$N_W = 5$]{\includegraphics[width=0.48\columnwidth]{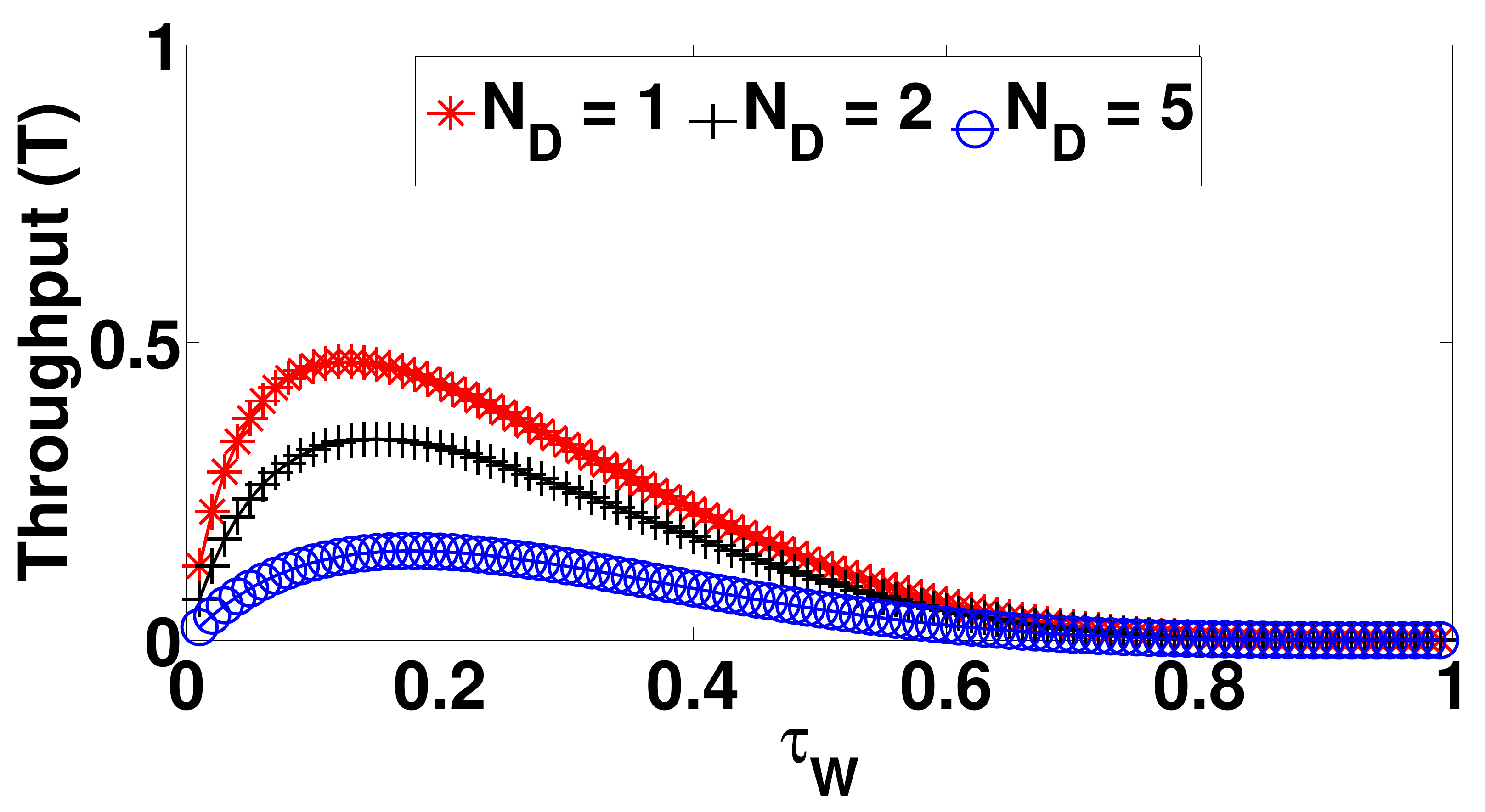}\label{fig:Thr3}}
  \caption{\small Impact of access probability on AoI and throughput. \SG{Figures~\ref{fig:Age1}-\ref{fig:Age3}} show access probability ($\tauD$) vs. AoI ($\Delta$) for \SnG{$N_D = 1$ and $N_D = 5$, respectively, for $N_W = 1, 2, 5$ and $\tauW = 0.2$}. \SG{Figures~\ref{fig:Thr1}-\ref{fig:Thr3}} show access probability ($\tauW$) vs. throughput ($T$) for \SnG{$N_W = 1$ and $N_W = 5$, respectively, for $N_D = 1, 2, 5$ and $\tauD = 0.2$}.}
  \label{fig:Age_Thr}
  \vspace{-5pt}
\end{figure}

\emph{Impact of access probability on average AoI and average throughput:}
Figure~\ref{fig:Age1} shows the AoI as a function of $\tau_D$ for a DSRC network that has a single node, for different numbers of nodes in the WiFi network. \SG{Nodes in }the WiFi network accesses the channel with probability $\tau_W=0.2$. The AoI achieved by the single node DSRC network increases with increase in the number of nodes in the WiFi network. This is because the larger the number of nodes in the WiFi network the smaller is the fraction of slots over which the DSRC network transmits successfully. Also, for any selection of size of WiFi network, the DSRC network sees AoI reduce as the chosen $\tau_D$ increases. More generally, as can be seen from~(\ref{Eq:ageutility}), when there is only a single DSRC node, AoI reduces with increasing $\tau_D$. Figure~\ref{fig:Age3} shows the AoI for when the DSRC network has $5$ nodes. A DSRC node must now suffer contention from other DSRC nodes too. This \emph{self-contention} penalizes the DSRC network for selection of large access probabilities. As is seen in the figure, the AoI is no longer monotonically decreasing. Finally, as was in the case of $N_D=1$, the presence of a larger WiFi network sees AoI increase for any selection of $\tau_D$.

In general, we observe that increased \emph{self-contention} (more DSRC nodes) makes larger $\tau_D$ less desirable. However, increased \emph{competition} (more WiFi nodes) doesn't have a similar impact on selection of $\tau_D$. Similar observations hold for the WiFi network. Figure~\ref{fig:Thr1} shows the throughput when there is just one WiFi node. The node sees its throughput reduce for larger $\tau_W$ across different numbers of nodes in the competing DSRC network. On the other hand when the number of nodes in the WiFi network is $5$, as seen in Figure~\ref{fig:Thr3}, it is more beneficial for the network to choose smaller $\tau_W$. Finally, larger numbers of competing nodes cause a drop in throughput achieved for any selection of $\tau_W$.

%

\begin{figure}[t!]
\vspace{-15pt}
  \centering
  \subfloat[$N_D = 1, N_W = 1$]{\includegraphics[width=0.48\columnwidth]{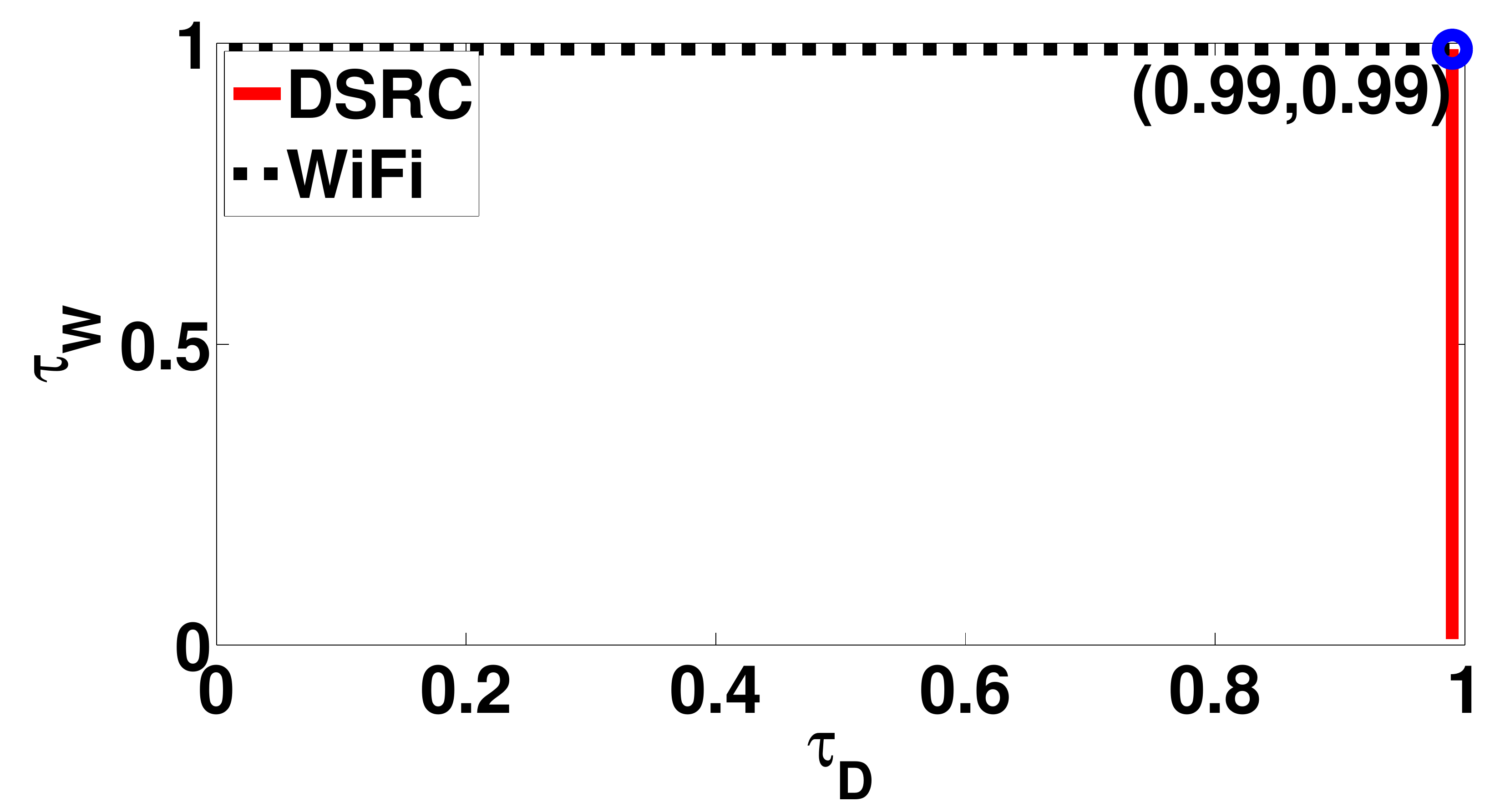}\label{fig:NE_nocost_11}}
  \enspace
  \subfloat[$N_D = 2, N_W = 2$]{\includegraphics[width=0.48\columnwidth]{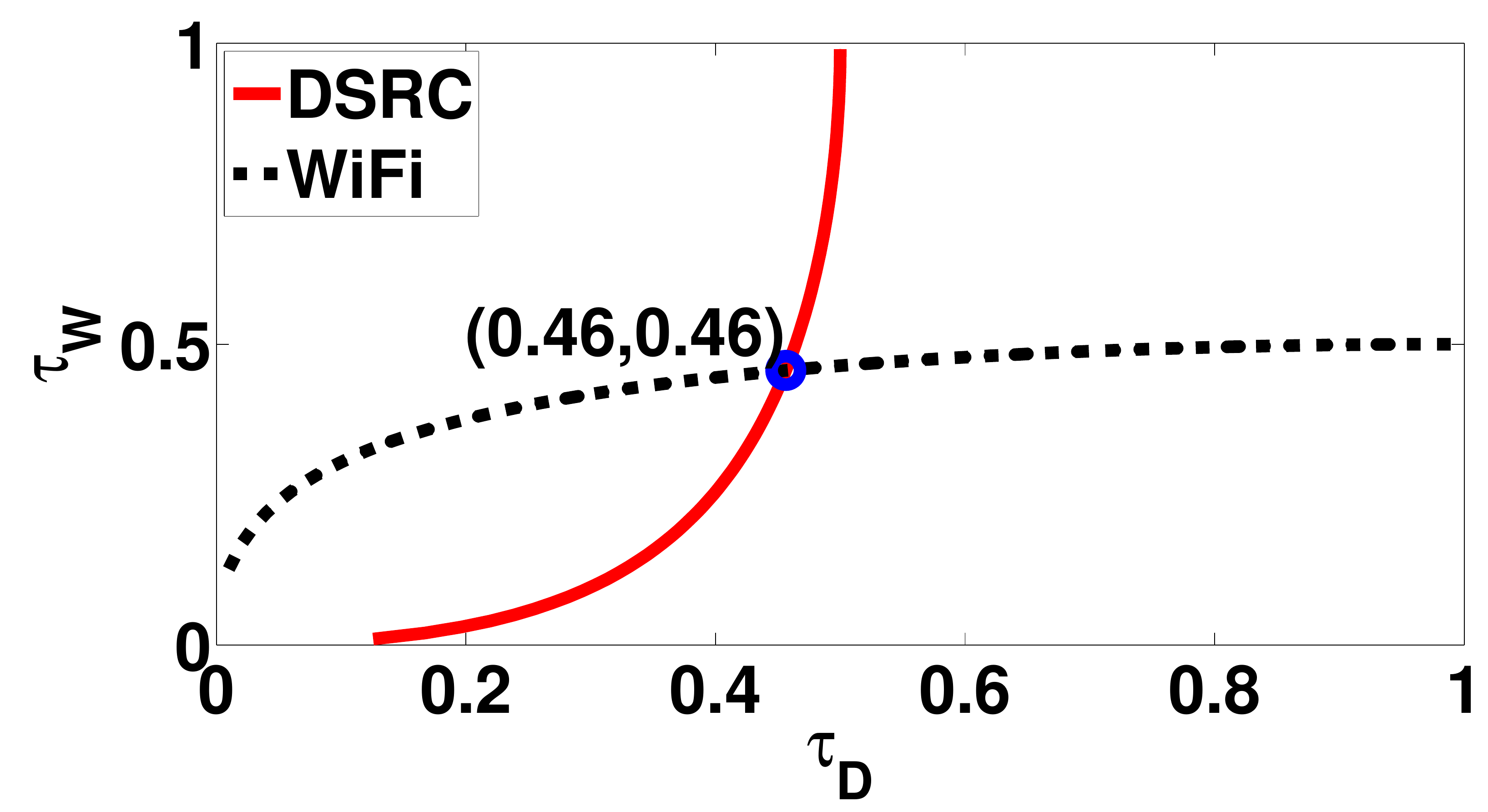}\label{fig:NE_nocost_22}}\\
  \subfloat[$N_D = 5, N_W = 5$]{\includegraphics[width=0.48\columnwidth]{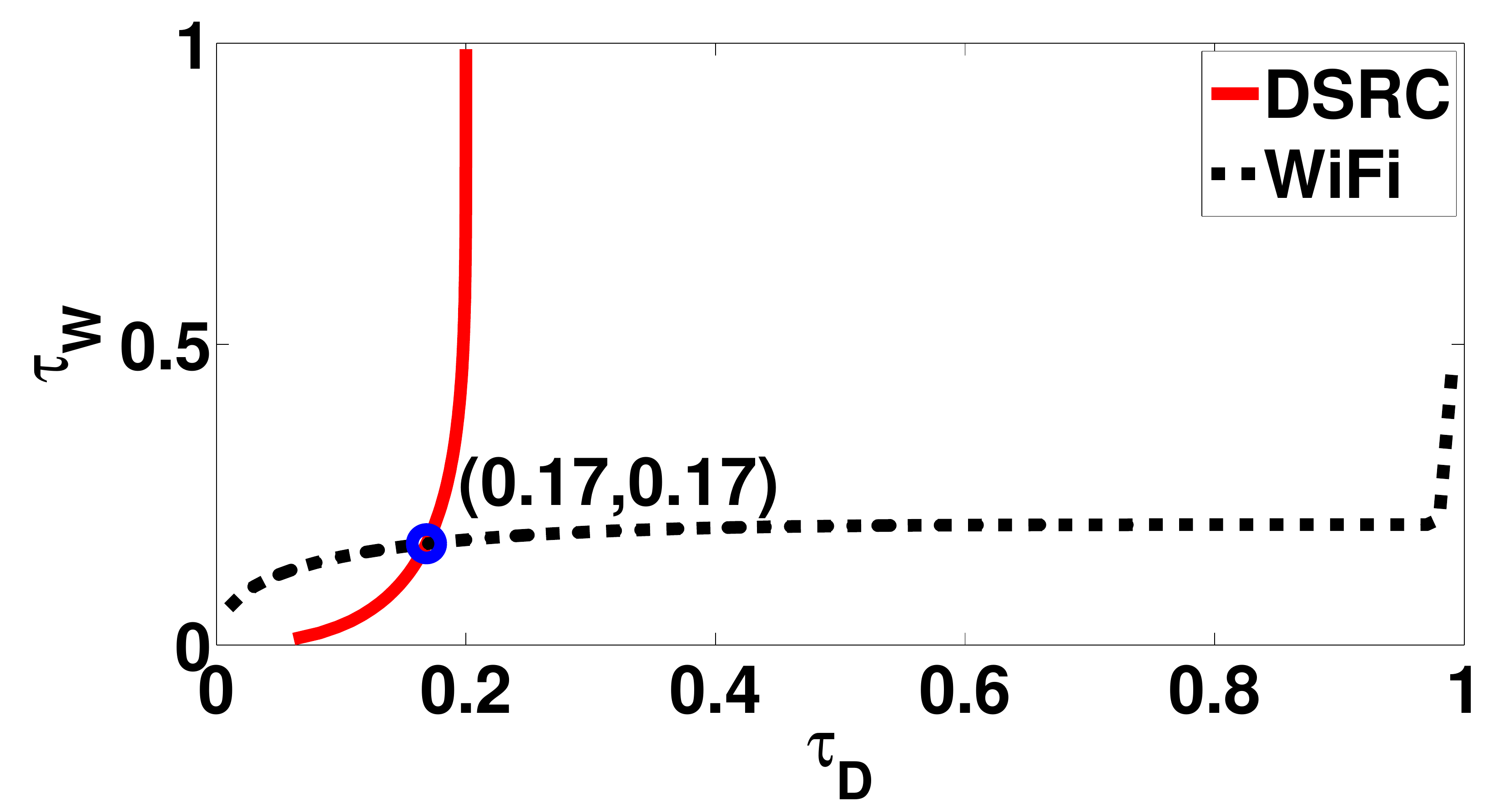}\label{fig:NE_nocost_55}}
  \enspace
  \subfloat[$N_D = 1, N_W = 1$]{\includegraphics[width=0.48\columnwidth]{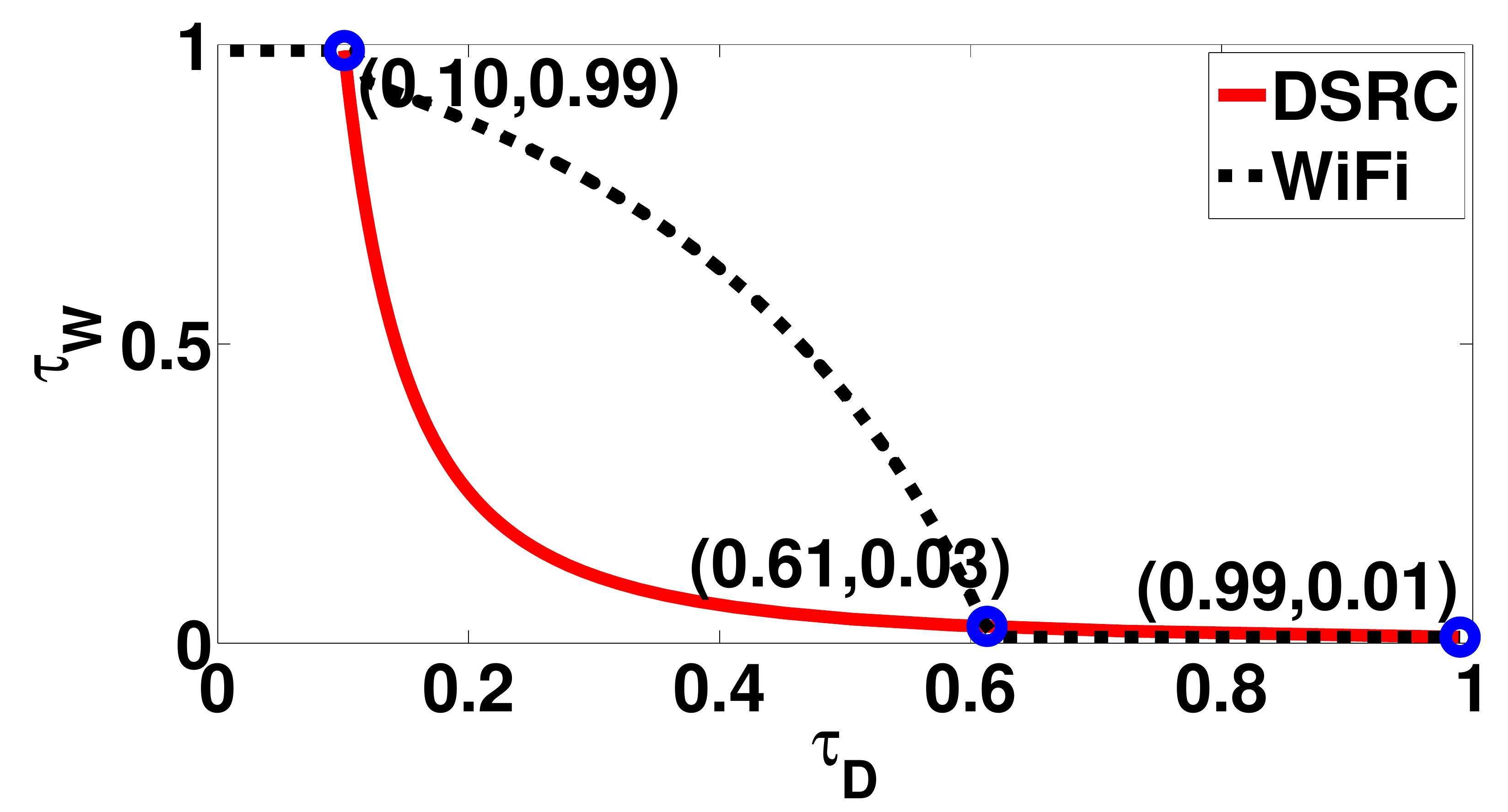}\label{fig:NE_cost_11}}\\
  \subfloat[$N_D = 2, N_W = 2$]{\includegraphics[width=0.48\columnwidth]{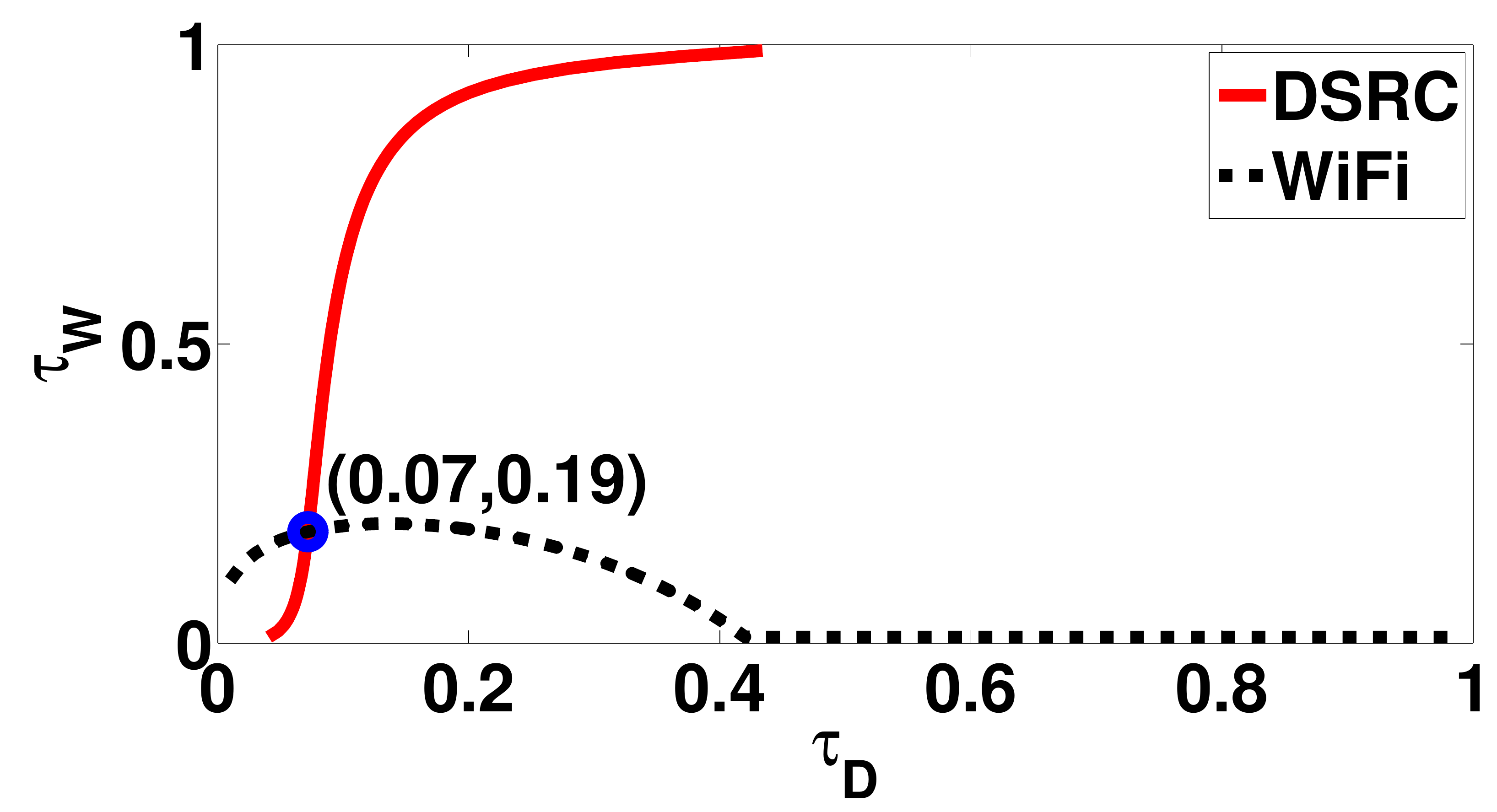}\label{fig:NE_cost_22}}
  \enspace
  \subfloat[$N_D = 5, N_W = 5$]{\includegraphics[width=0.48\columnwidth]{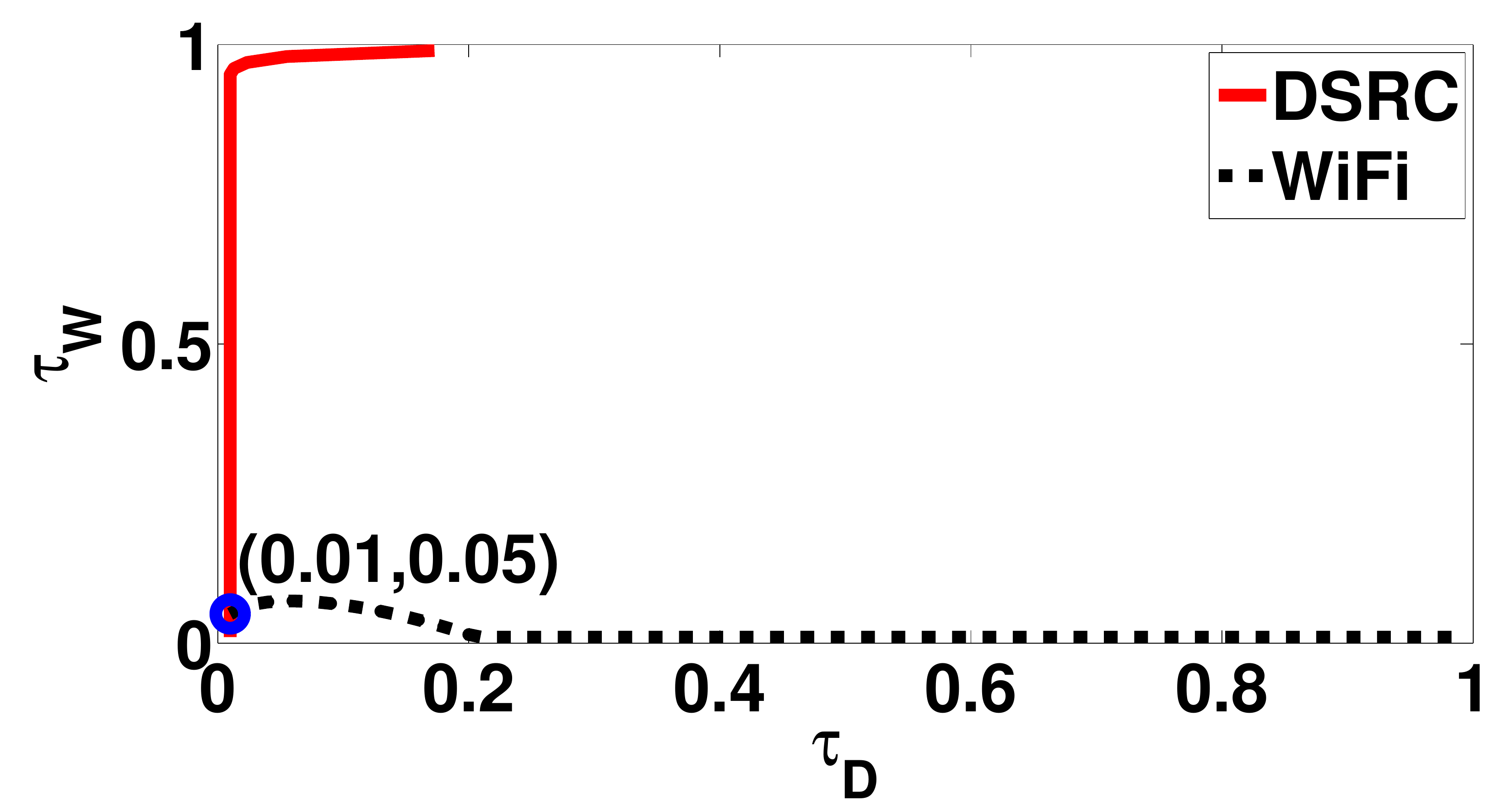}\label{fig:NE_cost_55}}
  \caption{\small Best response functions \SnG{for various} $N_D$ and $N_W$. For Figures~\ref{fig:NE_nocost_11}-\ref{fig:NE_nocost_55}, we set $w_{idle} = 0$ and $w_{col} = 0$. For Figures~\ref{fig:NE_cost_11}-\ref{fig:NE_cost_55}, we set $w_{idle} = \beta$ and $w_{col} = 1+\beta$, where $\beta = 0.001$.}
  \label{fig:NE}
\end{figure}

\emph{Nash equilibrium for the network coexistence game:} 
We start by considering the case when the networks don't pay any cost for wastage of spectrum, that is $w_{idle} = w_{col} = 0$. Figures~\ref{fig:NE_nocost_11}-\ref{fig:NE_nocost_55} shows the best response functions of the networks. The highlighted rows in Table~\ref{table:NE_nocost} show the corresponding AoI and throughputs. \SnG{The AoI (resp. throughput) of the DSRC (resp. WiFi) network increases (resp. decreases) as the number of nodes in the network increases for a fixed number of nodes of the other network.} Now consider the rows of Table~\ref{table:NE_nocost} that have $N_D=2$ DSRC nodes. As the numbers of WiFi nodes increase from $N_W=1$ to $N_W=5$, the access probability chosen by the DSRC network sees very little change. However, the access probability chosen by the WiFi network goes from a high of $0.99$ for $N_W=1$ to a low of $0.18$ for $N_W=5$. Now consider the rows that have $N_W=2$. One of the rows has $N_D = 2$ and the other has $N_D=5$. Again, the WiFi network's choice of access probability stays unaffected in comparison to a big change from $0.46$ to $0.17$ for the DSRC network. In summary, while \emph{self-contention} can lead to significant changes in choice of strategy, the strategy is relatively indifferent to changes in \emph{competition}.

Figures~\ref{fig:NE_cost_11}-\ref{fig:NE_cost_55} show the best response functions of the networks when a penalty is imposed for wastage of spectrum. We chose $w_{idle} = \beta$ and $w_{col} = 1+\beta$, where $\beta = 0.001$. Table~\ref{table:NE_cost} shows the NE and the corresponding AoI and throughput values. Since, $w_{idle}<w_{col}$, the cost of an idle slot is less than the cost of a collision, the players are less aggressive, their access probabilities are smaller, when compared to those for the no cost case tabulated in Table~\ref{table:NE_nocost}. Also, as in the absence of costs, unlike changes in \emph{competition}, changes in \emph{self-contention} lead to significant changes in the choice of access probabilities. Curiously, we obtain multiple NE for $N_D = 1$ and $N_W = 1$ (Figure~\ref{fig:NE_cost_11}). Observe that, among the three NE, the strategy $(0.61, 0.03)$ has both players do better.

\begin{table}[t!]
\footnotesize
\centering
\setlength\tabcolsep{8pt}
\begin{tabular}{|ccccc|}
\hline
$N_D$ & $N_W$ & ($\NED,\NEW$) & $\Delta(\NED,\NEW)$ & $T(\NED,\NEW)$\\
\hline
\textbf{1} & \textbf{1} & \textbf{(0.99, 0.99)} & \textbf{101.6015} & \textbf{0.0099}\\
2 & 1 & (0.50, 0.99) &  399.8980 & 0.2494 \\
\textbf{2} & \textbf{2} & \textbf{(0.46, 0.46)} & \textbf{12.9614} & \textbf{0.0803} \\
2 & 5 & (0.44, 0.18) & 9.9417 & 0.0288 \\
5 & 1 & (0.20, 0.99) & 1218.4  & 0.3268 \\
5 & 2 & (0.18, 0.44) & 35.2623 & 0.1060\\
\textbf{5} & \textbf{5} & \textbf{(0.17, 0.17)} & \textbf{26.8100} & \textbf{0.0380} \\
\hline
\end{tabular}
\caption{\small Nash Equilibrium, AoI and throughput values for varying $N_D$ and $N_W$ and $w_{idle} = 0$, $w_{col} = 0$ and $\beta = 0.001$. The highlighted rows correspond to Figures~\ref{fig:NE_nocost_11}-\ref{fig:NE_nocost_55}.}
\label{table:NE_nocost}
\vspace{-15pt}
\end{table}
\begin{table}[t!]
\footnotesize
\centering
\setlength\tabcolsep{8pt}
\begin{tabular}{|ccccc|}
\hline
$N_D$ & $N_W$ & ($\NED,\NEW$) & $\Delta(\NED,\NEW)$ & $T(\NED,\NEW)$\\
\hline
\textbf{1} & \textbf{1} & \textbf{(0.10, 0.99)} &  \textbf{983.0260} & \textbf{0.8981}  \\
  &   & \textbf{(0.61, 0.03)} & \textbf{1.5489} & \textbf{0.0174} \\
  &   & \textbf{(0.99, 0.01)} & \textbf{1.5117} & \textbf{0.0001} \\
2 & 1 & (0.07, 0.99) & 1436.3  & 0.8549  \\
\textbf{2} & \textbf{2} & \textbf{(0.07, 0.19)} & \textbf{10.2393} & \textbf{0.3035} \\
2 & 5 & (0.07, 0.06) & 8.6200 & 0.1115  \\
\textbf{5} & \textbf{5} & \textbf{(0.01, 0.05)} & \textbf{35.2475}  & \textbf{0.1460} \\
\hline
\end{tabular}
\caption{\small Nash Equilibrium, AoI and throughput values for varying values of $N_D$ and $N_W$ with $w_{idle} = \beta$, $w_{col} = 1+\beta$, where $\beta = 0.001$. The highlighted rows correspond to Figures~\ref{fig:NE_cost_11}-\ref{fig:NE_cost_55}.}
\label{table:NE_cost}
\vspace{-15pt}
\end{table}

\emph{Appropriate weight selection \SnG{may} lead to more efficient use of spectrum:}
Appropriate choice of weights $w_{idle}$ and $w_{col}$ \SnG{may} in fact nudge the networks to adopt strategies that lead to close to optimum sharing of the spectrum. We say that sharing is optimum when the AoI (resp. throughput) of a DSRC (resp. WiFi) node when using the equilibrium strategy is what it would be if one were to minimize AoI (resp. maximize throughput) of a network with the same number of total nodes, however, all of the DSRC (resp. WiFi) type. 

Table~\ref{table:optimal} shows the access probabilities $\OPTD$ and $\OPTW$ and the corresponding AoI and throughput values. Specifically, the first row that corresponds to $N=2$ shows the minimum age $\Delta(\OPTD)$ (resp. maximum throughput $T(\OPTW)$) that a DSRC node (resp. WiFi node) may achieve if only a $N=2$ node DSRC (resp. WiFi) network were to access the medium.
Table~\ref{table:NEWeights} shows a selection of weights, that nudges the networks closer to optimum sharing. Observe that selecting $w_{idle} = 0.001$ and $w_{col} = 150$, for $N_D = 1$ and $N_W = 1$, gives us AoI and throughput values that are close to the values in the row for $N=2$ in Table~\ref{table:optimal}. Similarly, optimum sharing is achieved for $N_D = N_W = 2$ and $N_D= N_W = 5$, by choosing $w_{idle} = 0.001$ and $w_{col} = 400$.
\begin{table}[t!]
\footnotesize
\centering
\begin{tabular}{|ccccc|}
\hline
$N$ & $\OPTD$ & $\Delta(\OPTD)$ & $\OPTW$ & $T(\OPTW)$\\[1mm]
\hline
\textbf{2} & \textbf{0.0268} & \textbf{2.5576} & \textbf{0.0306} & \textbf{0.4847}\\
4 & 0.0119 & 4.6505 & 0.0126 & 0.2407\\
10 & 0.0100 & 11.0723 & 0.0100 & 0.0946\\
\hline
\end{tabular}
\caption{\small \emph{Optimal} access probabilities $\OPTD$ (resp. $\OPTW$) and the corresponding AoI (resp. throughput) for nodes in a network that has only $N$ DSRC (resp. WiFi) nodes.}
\label{table:optimal}
\vspace{-0.5em}
\end{table}
\begin{table}[t!]
\footnotesize
\centering
\setlength\tabcolsep{4pt}
\begin{tabular}{|ccccccc|}
\hline
$N_D$ & $N_W$ & $w_{idle}$ & $w_{col}$ & ($\NED,\NEW$) & $\Delta(\NED,\NEW)$ & $T(\NED,\NEW)$\\
\hline 
1 & 1 & 0.001 & 150 & \textbf{(0.04, 0.04)} & \textbf{2.5793} & \textbf{0.4878}\\
  &   &       &     & (0.06, 0.02) & 1.8154 & 0.2163\\
  &   &       &     & (0.08, 0.01) & 1.6319 & 0.1003\\
2 & 2 & 0.001 & 400 & (0.01, 0.01) & 4.7179 & 0.2442\\
5 & 5 & 0.001 & 400 & (0.01, 0.01) & 11.0723 & 0.0946\\
\hline
\end{tabular}
\caption{\small Nash Equilibrium, AoI and throughput values obtained for different weight selections with the goal of nudging the networks to a close to \emph{optimum} (see Table~\ref{table:optimal}) use of spectrum.}
\label{table:NEWeights}
\vspace{-1em}
\end{table}

\emph{Stackelberg equilibrium for the network coexistence game:}
We solve for SE under two scenarios, (a) DSRC network is the leader and (b) WiFi network is the leader. Table~\ref{table:SE_nocost} and Table~\ref{table:SE_cost} show the SE and corresponding AoI and throughput values for different selections of $N_D$ and $N_W$ for the two scenarios, respectively, when (i) $w_{idle} = 0$ and $w_{col} = 0$, and (ii) $w_{idle} > 0$ and $w_{col} > 0$.

In Table~\ref{table:SE_nocost}, when the WiFi network is the leader and no spectrum wastage penalty is imposed, for $N_D = 2$ and $N_W = 2$, AoI at SE is $7.3323$ and the throughput is $0.0857$. On comparing these values with their corresponding NE counterparts in Table~\ref{table:NE_nocost}, i.e. AoI of $12.9614$ and throughput of $0.0803$, we see that while the leader sees a minor increase of $0.0054$ in throughput, the follower sees a considerable reduction of $5.6291$ in AoI. This is also seen for when $N_D = 5$ and $N_W = 5$. In these scenarios, there is a clear advantage to be a follower. The leader does at least as well as when using the NE strategy.

On the imposition of a spectrum wastage penalty, however, the leader may benefit greatly while the follower suffers. To exemplify, consider, in Table~\ref{table:SE_cost}, the SE strategy for $N_D = 5$ and $N_W = 5$, when DSRC is the leader. The AoI at SE is $9.5057$ and the throughput is $0.0043$. The corresponding values when using the NE strategy (Table~\ref{table:NE_cost}) are an AoI of $35.2475$ and a throughput of $0.1460$. This can be explained by a cost of idling that is much lower than that of collisions. It makes the follower's best response access probabilities small.
\begin{table}[t!]
\footnotesize
\centering
\setlength\tabcolsep{8pt}
\begin{tabular}{|ccccc|}
\hline
\multicolumn{5}{|c|}{\textbf{DSRC network as leader}}\\
\hline
$N_D$ & $N_W$ & ($\SED,\SEW$) & $\Delta(\SED,\SEW)$ & $T(\SED,\SEW)$\\
\hline
1 & 1 & (0.99, 0.99) & 101.6015 & 0.0099\\
2 & 2 & (0.32, 0.42) & 12.2014 & 0.1328 \\
5 & 5 & (0.10, 0.15) & 25.2029 & 0.0615 \\
\hline
\multicolumn{5}{|c|}{\textbf{WiFi network as leader}}\\
\hline
1 & 1 & (0.99, 0.99) & 101.5607 & 0.0099\\
2 & 2 & (0.41, 0.30) & 7.3323 & 0.0857\\
5 & 5 & (0.15, 0.10) & 16.7464 & 0.0405\\
\hline
\end{tabular}
\caption{\small Stackelberg equilibrium strategies, AoI and throughput values for (a) DSRC network is the leader and (b) WiFi network is the leader. We set $w_{idle} = 0$, $w_{col} = 0$ and $\beta = 0.001$.}
\label{table:SE_nocost}
\vspace{-10pt}
\end{table}

\begin{table}[t!]
\footnotesize
\centering
\setlength\tabcolsep{8pt}
\begin{tabular}{|ccccc|}
\hline
\multicolumn{5}{|c|}{\textbf{DSRC network as leader}}\\
\hline
$N_D$ & $N_W$ & ($\SED,\SEW$) & $\Delta(\SED,\SEW)$ & $T(\SED,\SEW)$ \\
\hline
1 & 1 & (0.61, 0.03) & 1.5489  &  0.0174\\
2 & 2 & (0.43, 0.01) & 3.2848 & 0.0049 \\
5 & 5 & (0.21, 0.01) & 9.5057 & 0.0043 \\
\hline
\multicolumn{5}{|c|}{\textbf{WiFi network as leader}}\\
\hline
1 & 1 & (0.10, 0.99) & 983.0260 & 0.8981 \\
2 & 2 & (0.07, 0.19) & 10.2393  & 0.3035 \\
5 & 5 & (0.01, 0.05) & 35.2475 & 0.1460 \\
\hline
\end{tabular}
\caption{\small Stackelberg equilibrium strategies, AoI and throughput values for (a) DSRC network is the leader and (b) WiFi network is the leader for $w_{idle} = \beta$, $w_{col} = 1+\beta$, where $\beta = 0.001$.}
\label{table:SE_cost}
\vspace{-10pt}
\end{table}

\section{Conclusion}
\label{sec:conclusion}
We formulated a strategic form game with DSRC and WiFi networks as players. The DSRC network desires a small age of information while the WiFi network desires a large throughput. Each network chooses the probability with which its nodes access the medium. We demonstrated Nash and Stackelberg equilibrium strategies for the game.

\section*{Acknowledgement}
\label{sec:ack}
This work has been partly supported by DeitY, Government of India, under grant with Ref. No. R$-23011/1/2014$-BTD. We thank the reviewers for their comments and Dr. Shreemoy Mishra and Dr. Gaurav Arora for their feedback.
\begin{spacing}{0.92}
\bibliographystyle{IEEEtran}
\bibliography{references}
\end{spacing}

\appendices
\section{Quasi-concavity of the payoff of the DSRC node}
The payoff $u_D(\tau_D,\tau_W)$ of the DSRC node is given by equation~(\ref{eq:agepayoff}). We need to show that the payoff is quasi-concave in $\tau_D$. This is equivalent to showing that $-u_D(\tau_D,\tau_W)$ is quasi-convex in $\tau_D$. Define $Q_W = (1 - \tau_W)^{N_W}$ and $Q'_W = N_W \tau_W (1 - \tau_W)^{(N_W - 1)}$. The derivative of $-u_D(\tau_D,\tau_W)$ with respect to $\tau_D$ is
\begin{align}
&\frac{\partial (-u_D(\tau_D,\tau_W))}{\partial\tau_D} = \alpha_1(\tau_D) + \alpha_2(\tau_D)+ \alpha_{col}(\tau_D)\nonumber \\ 
&\qquad\qquad\qquad\qquad\qquad- \alpha_{idle}(\tau_D),\nonumber\\
&\text{where}\nonumber\\
&\alpha_1(\tau_D) = \frac{\beta(1+\beta)}{2}\frac{Q_W N_D (1-\tau_D)^{N_D-1}}{(1 + \beta - (1 - \tau_D)^{N_D} Q_W)^2},\nonumber\\
&\alpha_2(\tau_D) = \frac{1}{\tau_D^2}\left(1 + \frac{(1+\beta)(N_D\tau_D - 1)}{Q_W(1 - \tau_D)^{N_D}}\right),\nonumber\\
&\alpha_{col}(\tau_D) = w_{col} (Q_W N_D (N_D - 1) \tau_D (1-\tau_D)^{N_D-2}\nonumber\\
&\qquad\qquad\qquad\quad + Q'_W N_D (1-\tau_D)^{N_D - 1}),\nonumber\\
&\alpha_{idle}(\tau_D) = w_{idle} Q_W N_D (1 - \tau_D)^{N_D - 1}.
\label{eqn:agePayoffDerivative}
\end{align}
To prove quasi-convexity of $-u_D(\tau_D,\tau_W)$ we will argue that as $\tau_D$ decreases from $1$ to $0$, the derivative $\frac{\partial (-u_D(\tau_D,\tau_W))}{\partial\tau_D}$ goes from being positive to negative, and no other changes in its sign take place. In other words, \SG{as $\tau_D$ goes from $0$ to $1$}, $-u_D(\tau_D,\tau_W)$ is monotonically decreasing in $\tau_D$ followed by being monotonically increasing.

We start by considering the case when $w_{idle}, w_{col} = 0$. The derivative is positive for all \SG{$\tau_D \in [1/N_D , 1)$} since both \SG{$\alpha_1(\tau_D)$ and $\alpha_2(\tau_D)$} are positive for such $\tau_D$.

For a certain $0 < \tau_D < 1/N_D$, $\alpha_2(\tau_D)$ becomes negative and stays negative as $\tau_D \to 0$. Further, as $\tau_D \to 0$, the absolute value of $\alpha_2(\tau_D)$ grows faster and dominates $\alpha_1(\tau_D)$. This is because, as \SG{$\tau_D$} becomes small, given that $0 < Q_W < 1$ and $0 < \beta < 1$, $1/\tau_D^2 > (1 - \tau_D)^{N_D - 1}/(1 + \beta - (1 - \tau_D)^{N_D} Q_W)^2$. As $\tau_D \to 0$, $\alpha_2(\tau_D)$ becomes negative, eventually the derivative becomes negative, and given the faster growth of the absolute value of $\alpha_2(\tau_D)$, the derivative stays negative.

Now consider the case when $w_{idle}, w_{col} > 0$. Note that if $w_{idle}$ is small enough to not make the derivative $\frac{\partial (-u_D(\tau_D,\tau_W))}{\partial\tau_D}$ in~(\ref{eqn:agePayoffDerivative}) negative as $\tau_D$ goes from $1$ to $0$, then quasi-convexity of $-u_D(\tau_D,\tau_W)$ follows as for the case of $w_{idle}, w_{col} = 0$. 

Consider the case when $w_{idle}$ is large and a large $\alpha_{idle}(\tau_D)$ causes the derivative to become negative at some \SG{$\tau_D = \hat{\tau}$} before the rapid reduction in $\alpha_2(\tau_D)$ for small $\tau_D$ as $\tau_D \to 0$. It is sufficient to show that the derivative stays negative for all $\tau_D \in (0, \hat{\tau})$. Note that $\alpha_2(\tau_D)$ decreases monotonically as $\tau_D$ goes from $1$ to $0$ and hence cannot cause the derivative to change sign as $\tau_D \to 0$. The term $\alpha_{col}(\tau_D)$ corresponding to collision costs becomes larger as $\tau_D \to 0$. However, the derivative of $\alpha_{col}(\tau_D)/\alpha_{idle}(\tau_D)$ is positive for all $0 \le \tau_D < 1$. Thus the ratio decreases as $\tau_D \to 0$ and $\alpha_{idle}(\tau_D) - \alpha_{col}(\tau_D)$ increases. This leaves the term $\alpha_1(\tau_D)$ for us to consider. We can rewrite the term as 
\begin{align}
\alpha_1(\tau_D) = \frac{Q_W N_D (1-\tau_D)^{N_D-1}}{\frac{2(1+\beta)}{\beta} \left(1 - \frac{(1-\tau_D)^{N_D} Q_W}{1+\beta}\right)^2}.
\label{eqn:alpha1a}
\end{align}
Comparing the numerator with $\alpha_{idle}$ in~(\ref{eqn:agePayoffDerivative}), it is clear that the increase in the denominator in~(\ref{eqn:alpha1a}) as $\tau_D \to 0$ will determine whether $\alpha_1(\tau_D)$ can change the derivative to positive before $\alpha_2(\tau_D)$ makes it negative. Let $\tau'$ be the value of $\tau_D$ at which the denominator of~(\ref{eqn:alpha1a}) becomes $1$, as $\tau_D$ goes from $1$ to $0$. Note that the denominator decreases monotonically as $\tau_D \to 0$. Let $\tilde{\tau}$ be the value of $\tau_D$ at which $\alpha_2(\tau_D)$ becomes zero. For $\tau_D$ smaller than $\tilde{\tau}$, $\alpha_2(\tau_D)$ will be negative. We have, using~(\ref{eqn:alpha1a}),
\begin{align}
\tau' &= 1 - \left(\left((1+\beta) - \sqrt{\frac{\beta(1+\beta)}{2}}\right)\frac{1}{Q_W}\right)^{1/N_D}\nonumber\\
&< 1 - \left((1+\beta) - \sqrt{\frac{\beta(1+\beta)}{2}}\right)^{1/N_D}.\label{eqn:tau_dash_ub}
\end{align}
Note that no such $\tau'>0$ may exist. The second inequality is obtained by observing that $0 < Q_W < 1$. The right-hand-side of the second inequality gives an upperbound on $\tau'$. Setting \SG{$\alpha_2(\tau_D) = 0$}, defined in~(\ref{eqn:agePayoffDerivative}), we can see that $\tilde{\tau}$ is obtained by solving 
\begin{align}
1 - N_D \tilde{\tau} = \frac{Q_W}{1+\beta} (1 - \tilde{\tau})^{N_D}.
\end{align}
The smallest value of $\tilde{\tau}$, for any given $\beta$ and $N_D$, is obtained for $Q_W=1$. It can be verified that this smallest value is larger than the upper bound on $\tau'$ given by~(\ref{eqn:tau_dash_ub}). This implies that the denominator of $\alpha_1(\tau_D)$ becomes smaller than $1$, if at all, only after $\alpha_2(\tau_D)$ becomes negative. The fact that $\alpha_2(\tau_D)$ grows negative faster than $\alpha_1(\tau_D)$ grows positive implies the derivative stays negative over $(0, \hat{\tau})$.
\section{Quasi-concavity of the payoff of the WiFi node}
We want to show that $-u_W(\tauD,\tauW)$ is quasi-convex. Define $Q_D = (1 - \tauD)^{N_D}$ and $Q'_D = N_D \tauD (1 - \tauD)^{(N_D - 1)}$. The derivative $\frac{\partial (-u_W(\tauD,\tauW))}{\partial \tauW}$ is given by
\begin{align}
&\frac{\partial (-u_W(\tauD,\tauW))}{\partial \tauW} = \alpha(\tauW) + \alpha_{col}(\tauW) - \alpha_{idle}(\tauW),\nonumber\\
&\text{where,}\nonumber\\
&\alpha(\tauW) = Q_D (1+\beta) (1-\tauW)^{N_W - 2}\nonumber\\ 
&\qquad\qquad\quad\biggl(\frac{Q_D (1 - \tauW)^{N_W}+(1 + \beta) (\tauW N_W - 1)}{(1 - Q_D (1 - \tauW)^{N_W}+ \beta)^2}\biggr),\nonumber\\
&\alpha_{col}(\tauW) = w_{col} (Q_D N_W (N_W - 1) \tauW (1 - \tauW)^{N_W - 2}\nonumber\\
&\qquad\qquad\qquad\quad + Q'_D N_W (1 - \tauW)^{N_W - 1}\SG{)},\nonumber\\
&\alpha_{idle}(\tauW) = w_{idle} Q_D N_W (1 - \tauW)^{N_W - 1}.
\label{eqn:thrPayoffDerivative}
\end{align}

It is a fact that $\alpha(\tauW)/\alpha_{idle}(\tauW)$ is monotonically increasing \SG{as $\tauW$ goes from $0$ to $1$}. The same is also true for $\alpha_{col}(\tauW)/\alpha_{idle}(\tauW)$. Also, \SG{as $\tauW$ goes from $0$ to $1$}, $\alpha(\tauW)$ monotonically increases, while $\alpha_{col}(\tauW)$ and $\alpha_{idle}(\tauW)$ monotonically decreases.

Consider $\tauW \in \SG{[1/N_W,1)}$. Note that $\alpha(\tauW)$, $\alpha_{col}(\tauW)$, and $\alpha_{idle}(\tauW)$ are all positive. The partial derivative may become negative at a certain $\tilde{\tau} \in \SG{[1/N_W,1)}$ in case $\alpha_{idle}$ is large. If so, then the derivative stays negative over all $\tauW \in (0,\tilde{\tau})$. This is simply because, as mentioned above, $\alpha(\tauW)/\alpha_{idle}(\tauW)$ and $\alpha_{col}(\tauW)/\alpha_{idle}(\tauW)$ are monotonically increasing in $\tauW$. So $\alpha_{idle}(\tauW) - \alpha(\tauW)$ and $\alpha_{idle}(\tauW) - \alpha_{col}(\tauW)$ increase as $\tauW \to 0$.

Now consider the case when $\tauW \in (0,1/N_W)$ and $\alpha_{idle}(\tauW)$ stays smaller than $\alpha(\tauW) + \alpha_{col}(\tauW)$ for  $\tauW \in (1/N_W,1)$. In this case, the derivative in~(\ref{eqn:thrPayoffDerivative}) may stay positive for all $\tauW \in (0,1/N_W)$ or may become negative at a certain $\tilde{\tau} \in (0,1/N_W)$. \SG{Need clarification:} One way the latter could happen is due to \SG{$\alpha(\tauW)$} becoming negative and in turn causing  $\SG{\alpha(\tauW)} + \alpha_{col}(\tauW)$ to become smaller than $\alpha_{idle}(\tauW)$. As explained earlier, if $\alpha_{idle}(\tilde{\tau})$ exceeds $\alpha(\tilde{\tau}) + \alpha_{col}(\tilde{\tau})$, $\alpha_{idle}(\tau_W)$ is greater than $\alpha(\tauW) + \alpha_{col}(\tauW)$ for all $\tau_W\in (0,\tilde{\tau})$. 

In summary, the derivative in~(\ref{eqn:thrPayoffDerivative}) either stays positive over all $\tauW$ or it goes from positive to negative as $\tauW$ decreases from $1$ to $0$. Thus, $-u_W(\tau_D,\tau_W)$ is quasi-convex.

\end{document}